# Data Transfer and Network Services management for Domain Science Workflows

Tom Lehman, Xi Yang, Chin Guok, Lawrence Berkeley National Laboratory, Berkeley, CA, USA

Frank Wuerthwein, Igor Sfiligoi, John Graham, Aashay Arora, Dima Mishin, Diego Davila, Jonathan Guiang, Tom Hutton, University of California San Diego, San Diego, California, USA

Harvey Newman, Justas Balcas, Caltech, Pasadena, CA, USA

## Abstract

This paper describes a vision and work in progress to elevate network resources and data transfer management to the same level as compute and storage in the context of services access, scheduling, life cycle management, and orchestration.  While domain science workflows often include active compute resource allocation and management, the data transfers and associated network resource coordination is not handled in a similar manner.  As a result data transfers can introduce a degree of uncertainty in workflow operations, and the associated lack of network information does not allow for either the workflow operations or the network use to be optimized.  The net result is that domain science workflow processes are forced to view the network as an opaque infrastructure into which they inject data and hope that it emerges at the destination with an acceptable Quality of Experience. There is little ability for applications to interact with the network to exchange information, negotiate performance parameters, discover expected performance metrics, or receive status/troubleshooting information in real time. Developing mechanisms to allow an application workflow to obtain information regarding the network services, capabilities, and options, to a degree similar to what is possible for compute resources is the primary motivation for this work. The initial focus is on the Open Science Grid (OSG)/Compact Muon Solenoid (CMS) Large Hadron Collider (LHC) workflows with Rucio/FTS/XRootD based data transfers and the interoperation with the ESnet SENSE (Software-Defined Network for End-to-end Networked Science at the Exascale) system.

## 1  Introduction

A key objective of this work is to develop mechanisms which facilitate the easy integration of advanced networks services into domain science workflows.  These workflows typically coordinate and orchestrate compute and storage resources, and initiate associated data management/movement activities.  High performance Research and Education (R&E) networks are a key infrastructure component needed for the data movement tasks.  Currently the domain science workflow processes are forced to view the network as an opaque infrastructure into which they inject data and hope that it emerges at the destination with an acceptable Quality of Experience.  There is little ability for applications to interact with the network to exchange information, negotiate performance parameters, discover expected performance metrics, or receive status/troubleshooting information in real time.  Developing mechanisms to allow an application workflow to obtain information regarding the network services, capabilities, and options, to a degree similar to what is possible for compute resources is the primary motivation for this work.  This specific work is focused on the Open Science Grid (OSG) / Compact Muon Solenoid (CMS) Large Hadron Collider (LHC) workflows with Rucio/FTS/XRootD based data transfers and the interoperation with the ESnet SENSE (Software-Defined Network for End-to-end Networked Science at the Exascale) system.  A brief description of each of these systems is provided below.

Open Science Grid (OSG) / Compact Muon Solenoid (CMS) workflows
The Open Science Grid (OSG) provides a distributed facility where the Consortium members provide guaranteed and opportunistic access to shared computing and storage resources.  The innovative aspects of the project are the maintenance and performance of a collaborative (shared & common) petascale national facility over hundreds of autonomous computing sites, for many thousands of users, managing data transfers, executing hundreds of thousands of jobs a day, and providing robust and usable resources for scientific groups of all types and sizes.  OSG operations include ongoing data distribution at aggregate of tens of terabytes a day between the distributed sites.

While the OSG compute resource management coordinates specific resources for individual users, the data transfers and associated network services do not.   As a result the data transfers can introduce a degree of uncertainty in workflow operations, and the associated lack of network information does not allow for either the workflow operations or the network use to be optimized.  The net result is that domain science applications and workflow processes are currently forced to view the network as an opaque infrastructure into which they inject data and hope that it emerges at the destination with an acceptable Quality of Experience. There is little ability for applications to interact with the network to exchange information, negotiate performance parameters, discover expected performance metrics, or receive status/troubleshooting information in real time.  Developing mechanisms to allow an application workflow to obtain information regarding the network services, capabilities, and options, to a degree similar to what is possible for compute resources is the primary motivation for this work.

The focus of this vision and design work is the OSG CMS LHC based workflows.  The OSG facilitates access to distributed high throughput computing for research in the US. The resources accessible through the OSG are contributed by the community, organized by the OSG, and governed by the OSG consortium.  There are many components of the OSG system including HTCondor, HPC schedulers, workflow managers, and others.

Rucio/FTS/XRootD based data transfers

Key components of this workflow include the Rucio system and File Transfer System (FTS) which are responsible for the data management and movement services supporting the CMS workflow operations. These systems are briefly described below:

- RUCIO - provides scientific collaborations with the functionality to organize, manage, and access their data at scale. The data can be distributed across heterogeneous data centers at widely distributed locations. The Rucio development team is located at CERN. Rucio is an open source software product external to this SOW. The fundamental concepts and architecture for the Rucio are described in a 2019 paper [1].

- File Transfer Service (FTS) - responsible for distributing LHC data across the Worldwide LHC Computing Grid. The FTS development team is located at CERN. FTS is an open source software product external to this SOW.

SENSE (Software-Defined Network for End-to-end Networked Science at the Exascale) system
The SENSE provides the mechanisms to enable multi-domain orchestration for a wide variety of network and other cyberinfrastructure resources in a highly customized manner. These orchestrated services can be customized for individual domain science workflow systems and requirements. SENSE services include Layer 2 Point to Point Network Connections, Layer 2 Multipoint Network Topologies, and Layer 3 Routed/Virtual Private Network (VPN) services. The SENSE system also provides a variety of interactive services allowing application workflows to ask open-ended questions about capabilities, negotiate with the networked infrastructure, or request network services in a highly abstract and workflow centric manner. The fundamental concepts and architecture for the SENSE are described in a 2020 paper [2].

Motivated by the above objectives, and building on the specific components described, the ongoing system design and prototype implementation activities are described in this paper. This system has the following key functions:

- Features which allow CMS and/or Rucio workflow processes to identify specific "dataflows" as needing "priority data movement" services.
- A network service which can provide "priority network use" across the full "end-to-end network path".
- Application Program Interface (API) driven interactions between Rucio and SENSE such that the priority data transfer management and network services can be coordinated
- An ability for CMS and/or Rucio workflows to change the priority of a specific dataflow during the full lifecycle of a data movement operation.

Further description of the quoted items above is provided below. This expands on their meaning in the context of this work, system vision, and prototype implementation:

- A "dataflow" is defined as a list of files that need to be moved from Site A to Site B. The key item to note here is the expectation that CMS and/or Rucio will need to define which files need priority transfer services, and be able to aggregate them on a site pair basis.

This definition of a "dataflow" is expected to change over time based on more complex site to site data movement requirements.
- A "priority data movement" uses the priority mechanism available in the existing Rucio system. The definition of "priority data movement" in this context is expected to change over time, as other policies or data may be used to define a specific dataflow priority.
- An "end-to-end network path" is defined as the full network path from the end systems at Site A to the end systems at Site B. A key item to note here is that this end-to-end path includes the wide area and site local network elements, along with the networking stacks inside the end systems.
- The term "priority network use" is defined as the assurance that the identified dataflow will have access to a minimum amount of network resources (i.e. bandwidth) across the end-to-end network path.

The following sections describe the key considerations from the CMS Workflow (Section 2) and the Rucio/FTS System (Section 3). The system design and prototype implementation plan, including the SENSE functions are also described (Section 4).

## 2 CMS Workflow Considerations

For this system vision and design, a set a workflow assumptions provide a starting point for the design activities:

- The CMS and Rucio components will decide which data needs to be transferred between two sites.
- Rucio and SENSE will interact to coordinate the provision of end-to-end network services and the initiation of the data transfer via the standard Rucio to FTS to XRootD methods.
- These layers above SENSE, i.e. CMS and Rucio, commit to fixed volumes that are known, and ready to be transferred at the time of the request to SENSE. i.e. whatever software makes the request to SENSE knows the exact list of files, and their sizes, that are meant to be transferred via this request. And that list of files is "ready" in a storage buffer that is accessible to the network with known performance characteristics that exceed the performance requested from the networks via SENSE.

To be explicit, any performance issues at the endpoints, both source and destination are considered error conditions, and commented on in the appropriate section below. The behavior designed for is allowed to make the assumption that performance issues at source and destination are rare, and can thus be treated as errors, rather than fundamentally being accounted for by the primary behavior designed into the system. This implies that SENSE calls are made only AFTER data is retrieved from tape, and resident on disk in the output buffer designed to be highly performant to support the requested network bandwidth.

Similarly, the upper layers above SENSE will have committed to source and destination before SENSE is invoked. This is important to spell out explicitly as it is not the default behavior of many of the currently used infrastructure software systems. They tend to specify destinations early, and pick sources late, or even dynamically adjust between different sources, keeping "second choices" active, and exercised to be ready to switch as necessary to achieve better performance. We defer any such more dynamic adjustments to future work, and beyond scope of the present work. Additional information regarding LHC workflow requirements and operations planning is available in the various study documents as noted in references [3] and [4].

## 2.1 Discussion of Scale

For this work it is assumed that in some future R&E networks, such as ESnet6, up to 80% of the bandwidth across the production infrastructure is available to be controlled via SENSE. That is, there will be a set of appropriately authorized organizations and individuals who will be able to control their use of the ESnet6 production infrastructure at up to 80% of the total available. The remaining 20% will always be reserved to unmanaged transfers as done today.

## 2.2 Application of Concepts and Scale to CMS:

The scale of computing for CMS in the High Luminosity (HL) LHC era is completely dominated by primary processing, and the movement of data to support it. This is true for expected CPU consumption, expected data volumes, and expected archival needs. There is some lack of clarity today whether or not it is also true for disk deployments, mostly because the error bars on how disk will be used are still very large. However, to the best of our knowledge today, we can assume that all large volume transfers between sources and destinations will be predictable and can be managed as bulk transfers are within this scope reduction.

We thus think that the scope definition expressed in the section "Discussion of Concepts" still covers more than 80% of the total transfer volume, and is thus consistent with the "Discussion of Scale". As we get closer to HL-LHC operations, we may have to revise these statements, and thus future additional scope may be necessary. However, for now, this is good enough.

It is worth clarifying some basic scale issues of CMS' science at the HL-LHC as follows:

1. The instrument produces collision data with 6.5MB per collision. During LHC runs (duty cycle ~ 30% during years the LHC runs. Running periods are 2-3 years long, with 1-2 year downtimes for maintenance in between) an average of 7.5kHz of data taking is assumed as target. The instrument thus produces ~ 50GBytes/sec of RAW data when the LHC is on, and integrates this to 364PBytes during each year the LHC is on. There are no realtime requirements on this data. The data is expected to be transferred to 7 tape archives distributed worldwide within a couple days of data taking, with the US archive at FNAL being the largest, expected to hold 40% of the data.

    a. From this we realize that the relevant timescales are ~ 1 day, and the relevant volume to move in that day is ~ 5PB leaving CERN. It's probably reasonable to ask for a few times this volume as capability per day to allow for catching up in data volumes from one day to the next. As the US is expected to receive 40% of this, it implies ~ 20GBytes/sec on average, or ~ 200Gbps CERN to FNAL on

average. This implies peaks of up to 1 Tbps or so, for hours at a time. **There are no relevant timescales smaller than an hour.**

## 2.3 Analysis Facility and Related Infrastructure and Services

CMS expects a transition towards so-called "Analysis Facilities". A science workflow for analysis has been proposed that would make transfers that are unpredictable today, due to the very nature of the caching and data federation, more predictable in the future.

The relevant service is called "ServiceX" and as of right now it's still in its very early development cycle. This service will not be addressed in this initial work. It is expected that these functions will be considered later as part of the larger consideration of the ongoing work with Data Lake and caching concepts.

## 3 Rucio, FTS, XRootD Based Storage System Key Considerations

This section describes the salient features of the system stack of Rucio, FTS, XRootD storage system design. This is the initial assessment of the key components as related to the design objectives of this project. It is expected that changes will be made as we learn more about the optimal way for the independent components to interact.

### 3.1 Data Management and Movement Workflow

Rucio is the overall data movement planner. It is assumed that Rucio will initiate interactions with SENSE for large bulk transfers between two sites. An end site is referred to as an "RSE" in Rucio, which stands for Remote Storage Element. Each RSE is configured in Rucio specifying:

- Storage endpoints, one for each protocol
  - Default port for each endpoint and protocol is specified
  - Namespace for endpoint is specified. This implies that an RSE can be composed of multiple independent filesystems, each serving a non-overlapping part of the total namespace, and served possibly via a different triplet of (endpoint, port, protocol).

- Preferences as to which protocol to use for what type of transfer for each endpoint (read/write/third party transfer).
  - Note that multiple endpoints can serve the same namespace via different protocols. E.g. in the example in the appendix we define a gsiftp and a webdavs endpoint, and define a preference for different types of access for the two protocols.

It is noted that the configuration allows for an RSE to be composed of multiple actual filesystems on independent hardware. i.e. the namespace and endpoint. Rucio will ask SENSE to create a large pipe between Site A and Site B. Rucio has the Sites A and B configured via their redirector, and protocol. SENSE returns back information sufficient for proper instructions to be forwarded to FTS. Rucio is responsible for translation from the Logical File Name (LFN) to the Physical File Name (PFN). Rucio then calls FTS with a prescription of the source and destination syntax, including preferred endpoint information, PFN in total at both source and destination. The Rucio daemon that does this is assumed to stay alive, and is capable of receiving callbacks from SENSE when SENSE wants to change strategy. SENSE will be responsible for integrating the site level data flows with the wide area network services.

### 3.2 Implications for Rucio and SENSE integration

There area few key observations to note regarding this integration between Rucio and SENSE as follows:

- The default data termination point information as configured via the RSE does not have to be the data termination point Rucio must use. That is, the aforementioned interaction will allow Rucio to tell FTS what data termination point to use.

- There may be timing issues with respect to when SENSE changes strategy. At that time, Rucio will generally already have told FTS to use the data termination point it was previously told by SENSE to use. The file transfers already handed from Rucio to FTS can thus no longer be changed in this model.

    - The objective is for Rucio to pass down to FTS only a small number of file transfer requests at a time, such that the number of unchangeable file transfers, i.e. pending in FTS, remains small.
    - Ideally SENSE will maintain the old path for some time, possibly at decreased bandwidth, so that the already committed transfers can complete. The alternative is that FTS transfers fails, retries multiple times, and then returns transfer failure to Rucio. We should not design a system that by design leads to transfer failures because those lead to unnecessary delays, and resource consumption inside FTS and Rucio to do the error handling.
    - Rucio may also need to tell SENSE when FTS is done with the old data termination point so that the old path can be completely disassembled.

FTS can be thought of as nothing more than a set of queues that queue up transfer requests, execute them, and handle the errors. FTS uses gfal-copy to support multiple protocols. So our working hypothesis is that there is nothing in FTS that is impacted in any way shape or form by the overall SENSE integration.

For the sake of being explicit, we describe here the structure of a Tier-2 that has an RSE entirely implemented using XRootD, and with a single filesystem under the hood. This implies an architecture that has a redirector as entrypoint, and N XRootD servers, where N is typically O(10).

For SENSE to define a path based on preferred data termination points of the O(10) XRootD servers it may be needed to think of the entire collection of redirectors and all XRootD servers as existing for each preferred data termination point on the same hardware. This could be implemented statically, i.e. preconfigured for a fixed range of ports.

This then becomes a design question where the preferred data termination point information for a given RSE is stored. SENSE will need to know, and possibly allocate based on this information, and the RSE will need to move data based on this information. The methods for keeping these systems and information synchronized is a design question. Evaluating how ATLAS and CMS keep track of this today may provide some insight.

## 4 SENSE Interoperation with CMS, Rucio, FTS, XRootD Based Workflows

Guided by the framework and objectives described in the previous sections, a system vision and design has been developed, and a prototype implementation is underway. This section provides a description of this design.

This system consists of the following key components:

- Rucio and SENSE Interoperation and Coordination
- Rucio and FTS Interaction
- FTS and Site DTN Cluster Interaction
- SENSE and Network Service Provisioning
- SENSE and Site Data Flow Integration with Network Services

Figure 1 shows an architecture and workflow diagram which includes each of these areas. Each of these components are described in more detail in the following section.

### 4.1 Rucio and SENSE Interoperation and Coordination

Rucio is the entity which has data location awareness and decides which data needs to be moved between sites in support of higher level LHC workflows. Rucio formulates detailed data movement commands and instructs the File Transfer System (FTS) to move the data. FTS acts as a data movement queue and interacts with the XRootD clusters at the sites to initiate and monitor data movement.

For this project we introduce an interaction between Rucio and SENSE to enable assignment and management of flow priority which will be reflected in network services provisioning. The objective is to provide Rucio with an ability to identify which flows should have higher priority in terms of network resources, and then adjust that over the lifecycle of the data movement operations.

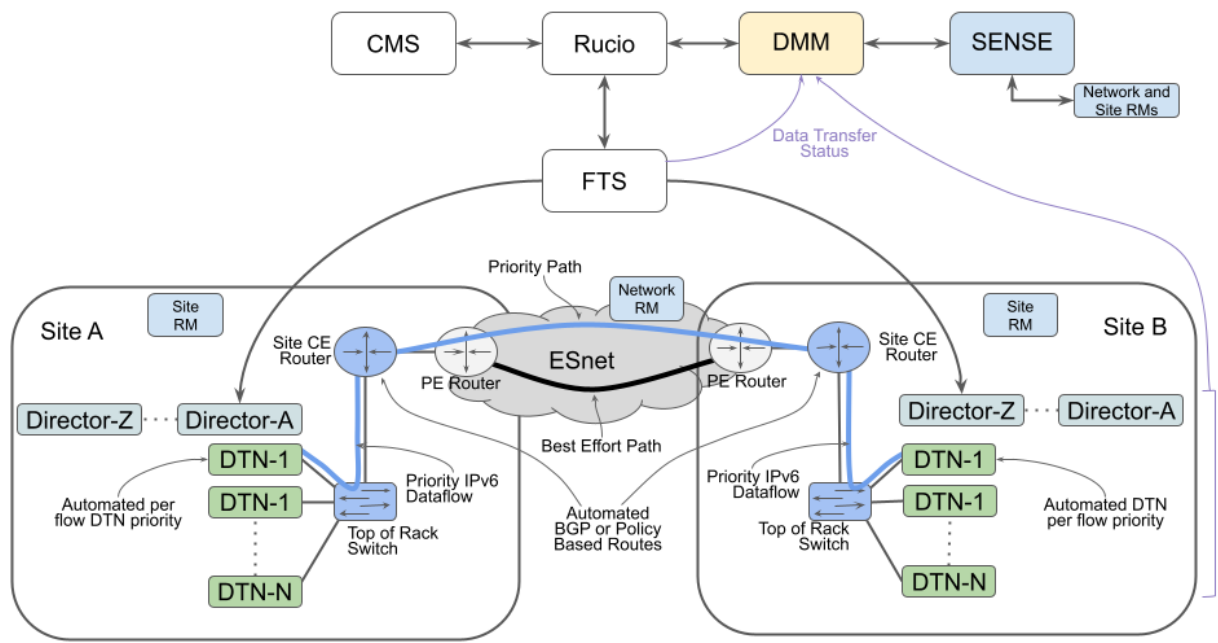

**Figure 1  SENSE and Rucio/FTS Interoperation Design**

A key objective of this early work is to minimize the impacts and changes needed to the current Rucio implementation and operations.  As a result, in this architecture a Data Movement Manager (DMM) has been introduced.  The DMM translates Rucio requests into more specific network services requests that the SENSE Orchestrator can act upon.  Future implementations may result in this DMM being integrated into Rucio, the SENSE Orchestrator, or remaining stand alone.  For this initial work Rucio sends a request to DMM which will include the following:

- data amount
- source site
- destination site
- relative priority

An assumption here is that data should be moved as soon as possible, with only the "relative priority" being factored into network resource allocation decisions.  That is, there is no concept of scheduling from Rucio's perspective for this initial implementation.  In addition, Rucio may send an update of the relative priority to DMS during the lifecycle of the data movement service.

A new method has been defined associating network services with specific data flows as indicated by Rucio.  This method is based on associating specific IPv6 subnets with specific priority network services.  The IPv6 subnets are uniquely associated with XRootD DTN cluster Directors.  Rucio will send messages to FTS based on this IPv6 Director information.  These associations can be adjusted over time as needed and requested by Rucio.  This use of IPv6 subnets is reflected in the response from the SENSE system, which includes information letting Rucio know which subnet to use for a specific service instance.  Figure 1 also shows feedback loops in the form of data transfer status from FTS and the site XRootD clusters.  This realtime data movement status information is future work and will not be included in the initial prototype implementation work.  This is intended to facilitate DMM and Rucio with making decisions during the lifecycle of a specific priority data movement service with respect to increasing or decreasing its priority level.  The SENSE system enables the adjustment of the priority, including placing the flow back into the best effort path, as requested by Rucio/DMM.

Section 4.4 provides additional information regarding how the data flows and network services are synchronized and integrated at the site level.

Figure 2 provides a description of the messaging workflow between the Rucio, DMM, and SENSE.  The workflow steps 1-5, and i in Figure 2 are related to this discussion in this section.

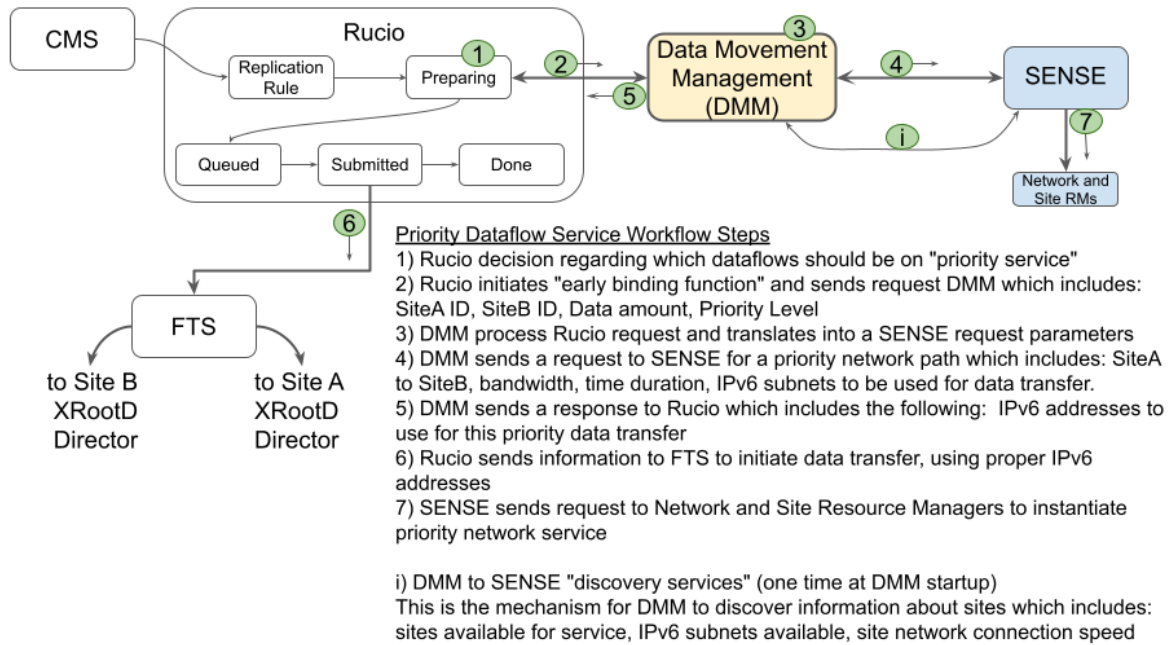

**Figure 2  SENSE and Rucio/FTS Interoperation Workflow**

## 4.2 Rucio and FTS Interaction and FTS to DTN Cluster Interaction

Rucio formulates detailed data movement commands and instructs the File Transfer System (FTS) to move the data.  FTS acts as a data movement queue and interacts with the XRootD clusters at the sites to initiate and monitor data movement.  For this project, there are no changes anticipated to the Rucio and FTS interactions.  The service priority mechanism will be realized via the Rucio to SENSE interaction which will result in specific IPv6 subnets and associated XRoodD directors.  The workflow step 6 in Figure 2 is related to this discussion in this section.

### 4.3 SENSE and WAN Network Service Provisioning
SENSE will utilize the information from the Rucio exchange to provision Wide Area Network (WAN) services.  These network services are constructed such that all network resources are available all the time.  That is, best effort traffic will be normally flowing over the WAN links, and when a priority service is provisioned, that will simply give the identified flow priority over the best effort traffic.  The SENSE architecture includes an Orchestrator function which will interact with a Network-Resource Manager to realize these wide area network provisioning actions. The workflow step 7 in Figure 2 is related to this discussion in this section.

### 4.4 SENSE and Site Data Flow Integration with Network Services
The objective for this area was to limit the impacts and changes needed to site deployments. However, this is an area where some additional new functions will be needed.   The intention for these changes are:

- Minimal changes to site deployments
- A preference for one time static network element configurations as opposed to many dynamic configurations
- A preference for techniques based on Layer 3 routing as opposed to Layer 2 path provisioning for site level flow management
- Do not require advanced features which may only be available with expensive routers/switches

The solution developed is as follows:

- The Site XRootD cluster will have multiple IPv6 addresses added to their dataplane interface. Each of these IPv6 addresses will be from a different subnet.
- There will be multiple XRootD directors, with each one responsible for a single IPv6 subnet.  In this manner, the Rucio to FTS message path can control which IPv6 subnet will be used for a specific flow.  This will also allow SENSE to integrate specific end site flows to specific network services.
- The SENSE/ESnet WAN network services can be terminated on the ESnet Provider Edge or the Site Customer Edge router.  Some dynamic configuration will be needed to integrate a specific site level flow (IPv6 subnet) to a specific WAN network priority service.
- This dynamic integration of a site flow to a WAN network priority service should be not data movement impacting, and can be flexibly adjusted over the lifecycle of a data movement channel.
- There are also some priority level services that will be implemented on the site network elements and the Data Transfer Nodes (DTNs).  The SENSE architecture includes an Orchestrator function which will interact with a Site-Resource Manager (Site-RM) to realize these site networking and DTN provisioning actions. These function are described below:

- Customer Edge or Provider Edge Router - The dataflows initiated via the standard FTS to XRootD mechanisms, will be started on the specific IPv6 dataplane addresses as agreed to between the earlier Rucion and SENSE Orchestrator interactions. The SENSE Site-RM will use this knowledge to dynamically configure policy based routes on the site customer edge or provider edge router to steer the priority data flow onto the proper wide area service.
- Top of Rack Switch - One time configurations on this switch to read the packet priority marking placed by the DTNs will allow an easy way to provide managed priority access to the egress queues of the top of rack switch.
- DTN Priority Service - The SENSE system, via the Site-RM, can manage resources utilization on the DTN nodes. This can include management of priority on network interfaces, other host resources, packet marking, and traffic shaping. For this system, the SENSE Site-RM will use its knowledge of which IPv6 address should be using the priority service and do the following:

  - use Linux traffic control features to provide specific priority to the appropriate dataflow
  - mark the dataflow packets such that the top of rack switch can identify and place priority packets in the proper egress queue

These types of features should not be data movement impacting, and can be flexibly adjusted over the lifecycle of a data movement channel, as instructed by Rucio.

## 5 Summary

This paper described an architecture and design for workflow systems, such as Rucio, to obtain a priority network service between any two sites of interest. Rucio can then use this network service to flexibly specify the associated dataflows which will use this priority service. This system also allows Rucio to engage in continuous adjustment of which dataflows use the priority network service, across the full lifecycle of a specific data transfer event, or across multiple data transfer events. This system does not require Rucio to know or process anything regarding network topologies or service details. In this design, Rucio is responsible for knowing which data sets need to be moved between which sites, and also being able to communicate a relative priority level to the SENSE system. In the future it may be helpful if systems such as CMS and/or Rucio could also define more deterministic set data transfer requirements, such as "this data set needs to be moved between Site A and Site B, anytime within the next 2 days". This would allow the SENSE system to more efficiently plan network utilization, and work toward a key goal of increasing overall network utilization, while improving user quality of experience at the same time.